\PassOptionsToPackage{hypcap=false}{caption}
\documentclass[10pt,twocolumn,letterpaper]{article}

\usepackage{cvpr}              
\definecolor{cvprblue}{rgb}{0.21,0.49,0.74}
\usepackage[pagebackref,breaklinks,colorlinks,allcolors=cvprblue]{hyperref}
\usepackage{xcolor}
\usepackage{listings}
\lstdefinestyle{plaintext}{
  language={},                
  basicstyle=\ttfamily\small,
  numbers=left,
  numberstyle=\tiny,
  frame=single,
  breaklines=true,
  columns=fullflexible,
  keepspaces=true,
  showstringspaces=false
}
\usepackage{graphicx}
\usepackage{cuted}    
\usepackage{caption}
\usepackage{xcolor}
\usepackage{fvextra}

\DefineVerbatimEnvironment{PromptText}{Verbatim}{
  fontsize=\scriptsize,
  breaklines=true,
  breakanywhere=true,
  breaksymbol={},
  breaksymbolleft={},
  breaksymbolindent=0pt,
  numbers=none,
  formatcom=\color{black},
  vspace=0pt
}
\DefineVerbatimEnvironment{PromptTextBlue}{Verbatim}{
  fontsize=\scriptsize,
  breaklines=true,
  breakanywhere=true,
  breaksymbol={},
  breaksymbolleft={},
  breaksymbolindent=0pt,
  numbers=none,
  formatcom=\color{blue},
  vspace=0pt
}

\usepackage{float}
\usepackage{url}

\def\paperID{6822} 
\def\confName{CVPR}
\def\confYear{2026}

\title{Acoustic Field Video for Multimodal Scene Understanding}

\author{Daehwa Kim\\
Carnegie Mellon University\\
Pittsburgh, Pennsylvania, USA\\
{\tt\small daehwak@cs.cmu.edu}
\and
Chris Harrison\\
Carnegie Mellon University\\
Pittsburgh, Pennsylvania, USA\\
{\tt\small chris.harrison@cs.cmu.edu}
}

\begin{document}
\maketitle

\begin{strip}
    \centering
    \vspace{-15mm}
    \includegraphics[width=1\linewidth]{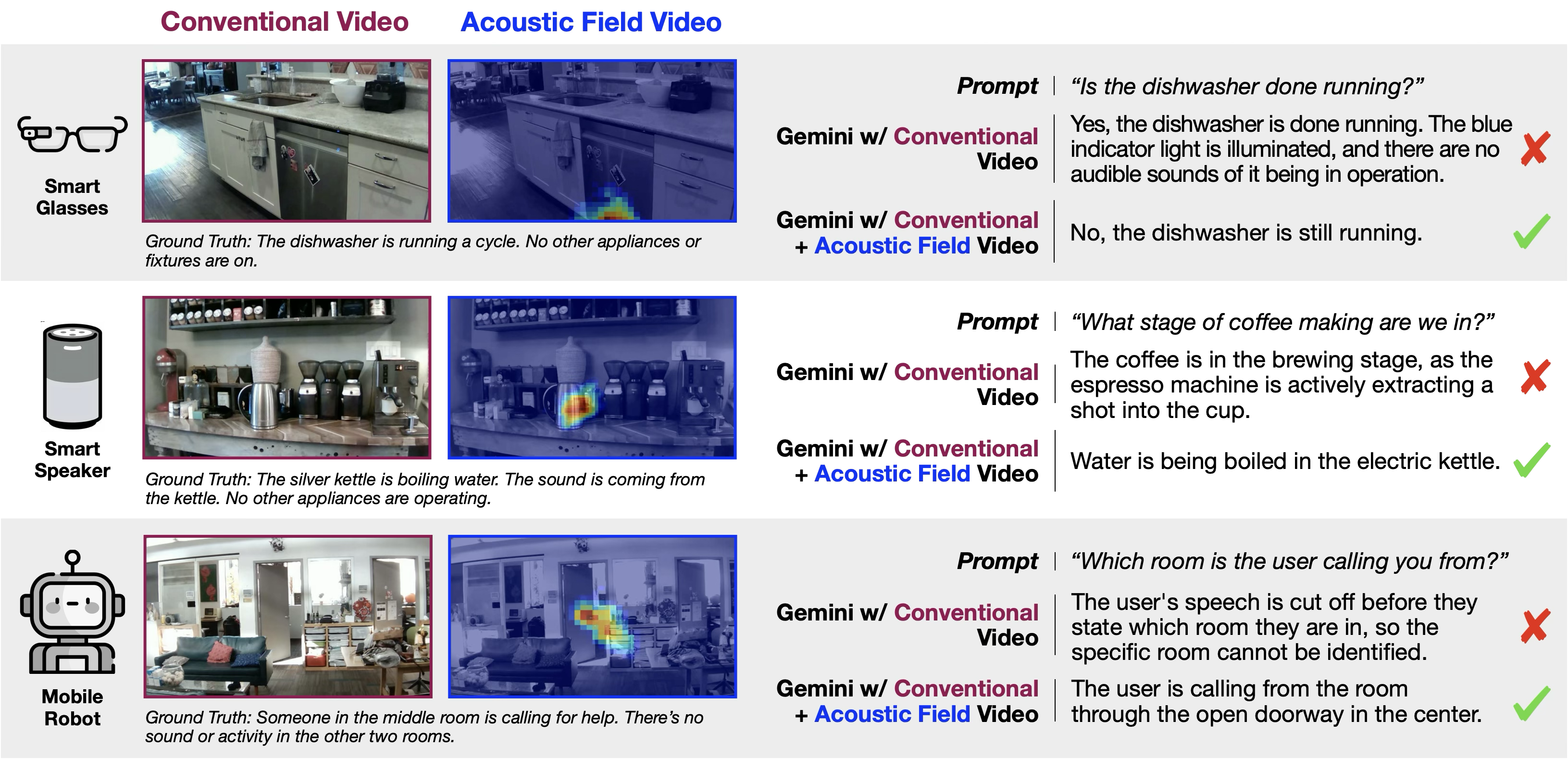}
    \captionof{figure}{Many everyday scene-understanding tasks remain underconstrained when relying solely on conventional video (with audio) input. We show that adding acoustic field video, a modality that visualizes the spatial distribution of sound, significantly improves multimodal reasoning. Moreover, the necessary hardware can be practically integrated into many platforms, from smart glasses to robots.} 
    \label{fig:teaser}
\end{strip}

\begin{abstract}
We introduce and explore a new multimodal input representation for vision–language models: acoustic field video. Unlike conventional video (RGB with stereo/mono audio), our video stream provides a spatially grounded visualization of sound intensity across a scene, offering a new and powerful dimension of perceptual understanding. Our real-time pipeline uses low-cost beamforming microphone arrays --- already common in smart speakers and increasingly present in robotics and XR headsets --- yet this sensing capability remains unutilized for scene understanding. To assess the value of spatial acoustic information, we constructed an evaluation set of 402 question–answer scenes, comparing a SOTA VLM given conventional video with and without paired acoustic field video. Results show a clear and consistent improvement when incorporating spatial acoustic data; the VLM we test jumps from 38.3\% correct to 67.4\%. Our findings highlight that many everyday scene understanding tasks remain underconstrained when relying solely on visual and audio input, and that acoustic field data provides a promising and practical direction for multimodal reasoning. A video demo is available at \href{https://daehwakim.com/seeingsound}{https://daehwakim.com/seeingsound}
\end{abstract}
\vspace{-3mm}
\section{Introduction}
\label{sec:intro}

Vision–language models (VLMs) demonstrate remarkable performance across a wide range of multimodal reasoning tasks, from visual question answering to embodied dialogue. However, despite their growing sophistication, most VLMs remain grounded in just two sensory modalities: visual imagery (RGB video) and non-spatial audio (typically stereo streams). This leaves a critical gap in perceptual understanding: the ability to reason about where, and thus from what objects, sounds are being generated within a scene.

Sounds in natural environments are often byproducts of physical processes and human activity: engines hum, electronics beep, fans blow, doors creak, and kitchen appliances whir. These distributed auditory cues accompany specific object states, motions, or interactions, providing valuable implicit information about the world. For example, a boiling kettle indicates temperature and phase change; a whirring fan reveals mechanical activity; muffled speech implies distance or occlusion. Sound, therefore, is not merely an auxiliary sensory channel, it is a diagnostic signal of a scene’s underlying dynamics. Localizing these acoustic signatures provides a powerful complementary signal to visual data, particularly for reasoning about state, activity, and causality.

For this reason, spatial hearing plays a central role in human perception, enabling us to locate sound sources, resolve ambiguities in complex environments, and integrate auditory cues with vision to infer scene structure. While recent multimodal research has incorporated audio spectrograms or raw waveforms, these representations encode frequency and time but lack explicit spatial grounding. As a result, current models struggle with questions requiring the linkage between sound and location. This is especially true for generic or repeated sounds, such as motors humming, fans blowing, or electronics beeping, which could originate from many objects within a scene. In such cases, audio alone cannot resolve the source (see Figures~\ref{fig:teaser}, \ref{fig:successful_examples} and \ref{fig:failure_attention}).

To address this gap, we introduce a new input modality for multimodal understanding: acoustic field video. Each frame encodes the spatial distribution of acoustic energy across a scene, effectively visualizing where sounds occur. Our real-time pipeline generates this stream using a microphone array and well-established beamforming algorithms, producing acoustic maps that are spatially and temporally aligned with RGB video frames and conventional audio. This alignment yields a rich, multimodal representation that unifies visual appearance, auditory content, and the spatial structure of scene sounds. While acoustic fields are not new (often referred to as sound pressure level (SPL) maps or related terms in acoustics), this is the first work to employ them as direct input to VLMs and to evaluate their utility for scene understanding and reasoning.

Our implementation requires low-cost microphone arrays, hardware already common in smart speakers and increasingly in domains such as robotics and XR headsets. Compared to other extended sensory modalities explored for use with VLMs --- most notably thermal imagery~\cite{Girdhar_2023_CVPR, Hwang_2015_CVPR} --- microphone arrays built from commodity MEMS components are inexpensive, small, and power efficient (e.g., the ICS-41350 MEMS microphone costs under \$0.75 in volume, measures 3.5×2.65×0.98 mm, and consumes just 185~$\mu$A in its always-on mode). This allows for microphone arrays to be integrated into even highly-constrained worn devices, such as Meta’s new Ray-Ban Display Glasses, which contain a six-microphone array.

To demonstrate the potential of this approach, we constructed a diagnostic benchmark of 402 question–answer (QA) pairs that span domestic, commercial, and industrial scenes. We open source both this dataset and our software pipeline. For evaluation, we use Gemini 2.5 Pro, a representative state-of-the-art VLM, to compare QA performance using traditional RGB+sound inputs against the same data augmented with acoustic field video. The addition of this modality yields substantive improvements (from 38.3\% to 67.4\% accuracy), particularly for questions involving localization, spatial attribution, and multi-source reasoning—tasks often unsolvable from RGB and stereo audio alone. 


\section{Related Work}

\subsection{Multimodal and Vision-Language Models}


Recent years have seen rapid progress in large-scale vision–language models (VLMs) that jointly process vision and text. Models such as Flamingo~\cite{alayrac2022flamingo}, Kosmos-2~\cite{peng2023kosmos2groundingmultimodallarge}, LLaVA~\cite{liu2023visual}, and Gemini learn cross-modal representations that support image captioning, visual QA, and grounded dialogue without task-specific supervision. Video-centric VLMs further extend this to temporal reasoning and audio for open-ended video QA~\cite{cheng2024videollama2,munasinghe2023pgvideollava}, and AVQA systems that incorporate mono-channel audio~\cite{li2023progressive,ye2024cat,lee2025bridging} further highlight the value of combining sound and vision for scene understanding.

While these models achieve impressive performance, they remain primarily bimodal, relying on static RGB images or video streams and text. Audio, when included, is typically represented as 1D waveform embeddings or spectrograms that encode temporal–frequency structure but lack explicit spatial grounding. Consequently, even advanced VLMs are limited in their ability to infer where in a scene a sound originates, a key factor in human spatial understanding and embodied perception. Our work complements this line of research by introducing a spatially grounded acoustic representation that can be seamlessly integrated into VLM pipelines, extending their perceptual range beyond visual and textual modalities.



\subsection{Modalities Beyond RGB Video \& Audio}

A parallel line of research explores how additional modalities—such as mono/stereo audio, depth, thermal, and motion—can complement visual and linguistic cues. AudioSet~\cite{audioset2017gemmeke} and its derivatives provide large-scale audio–visual corpora for cross-modal pretraining, while models such as SoundSpaces~\cite{chen2020soundspaces}, ImageBind~\cite{Girdhar_2023_CVPR}, LanguageBind~\cite{zhu2024languagebind}, and AVE~\cite{Tian_2018_ECCV} explore joint embeddings between text/vision to other modalities for localization, retrieval, and semantic understanding. These efforts have shown that sound enriches scene understanding by providing temporal continuity, semantic cues, and motion-related context not easily captured by vision alone.

However, most existing audio–visual–language models treat sound as a non-spatial signal—typically as mel-spectrograms or learned audio embeddings aggregated over time. This approach captures what is heard, but not where it occurs. As a result, models struggle with questions or reasoning tasks that depend on spatial localization, multi-source attribution, or occlusion relationships. In contrast, our proposed acoustic field (AF) video introduces an explicitly spatial acoustic modality, derived from beamformed microphone arrays, that visualizes sound energy across a scene. This enables multimodal models to reason about both the identity and spatial distribution of sound sources, bridging the gap between existing audio–visual fusion approaches and fully spatialized multimodal perception.

Methods such as Progressive Spatio-Temporal Perception for Audio-Visual Question Answering~\cite{li2023progressive} model temporal dynamics and cross-modal correlations to improve AVQA performance. Bridging Audio and Vision~\cite{lee2025bridging}, explores self-supervised methods to associate sounds with their visual sources in video. While these approaches successfully link objects and their auditory signatures, they treat audio as a non-spatial signal, limiting the ability to reason about precise sound locations or objects. As a result, if there are multiple objects in the scene of the same type, there can be confusion. Likewise, if a sound is generic, such as a motor hum, it may not be possible to guess which device is producing the sound in a complex scene such as a kitchen or workshop. The main example offered in \cite{lee2025bridging} is two clarinet players, which start at different times, and thus can rely on visuo-temporal data for disambiguation. But this crucial visual information is not readily available in a wide range of scenes, especially non-human scenes. When actions are not visible, such as a computer fan running at max speed in a server rack, or a leaking faucet or toilet, there will likely be confusion. 

In contrast, our approach leverages real-world live sensor data to derive spatial acoustic maps that explicitly encode the spatial distribution of acoustic energy, enabling multimodal models to disambiguate. To our knowledge, no prior work has explored using acoustic field video for audio-visual question answering.

\subsection{Audio Scene Understanding}
Beyond multimodal VLMs, a large body of work studies audio scene understanding without video. Audio-only approaches tackle acoustic scene classification, event detection, and audio-driven video understanding, including predicting visual motion or learning visual features from sound in a self-supervised way \cite{NIPS2016_7dcd340d, laput2018ubicoustics, Arandjelovic_2017_ICCV, afouras2020video, Zhou_2018_CVPR, GIRIN201953}. Audio QA benchmarks such as AVQA~\cite{Li_2022_CVPR} further require models to answer semantic questions about sound events and environments from spectrograms or waveforms. These methods show that sound alone carries rich information about activities, materials, and object states, but they operate in non-spatial feature spaces and are not designed to interface directly with general-purpose VLMs.

When video is available, most work emphasizes sound source localization and separation using vision. Early audio-visual correspondence methods such as “Look, Listen and Learn”~\cite{Arandjelovic_2017_ICCV} and “Sound Source Localization in the Wild”~\cite{Liu2022inthewild} localize sounding objects from unconstrained video, and later approaches refine this with contrastive learning and self-supervision for robust localization \cite{Sun_2023_CVPR, senocak2024aligning, mo2022localizing, liu2022exploiting, mo2022closerlook, park2023marginnce, liu2022exploiting, chen2021hardway, Sun_2023_CVPR, Zhao_2018_ECCV, jia2025seeingsoundhearingsight}. These pipelines typically output a heatmap or mask used for detection, retrieval, or sound isolation, rather than as a primary input to a reasoning system. In contrast, we derive real-time acoustic field (AF) video from beamformed microphone arrays and feed this spatially explicit sound-intensity map directly into an off-the-shelf VLM alongside RGB video, allowing us to study how explicitly spatialized acoustics improve zero-shot scene understanding and question answering rather than localization or separation alone.

\section{Implementation}
\label{sec:implementation}
We now describe the main components of our system, including both hardware and software. Our pipeline is made open source at https://www.github.com/anonyimized-for-review. 

\subsection{Hardware}
\label{sec:hardware}
We note that many contemporary devices already contain microphone arrays. For instance, the Apple HomePod includes a six-microphone array~\cite{ifixit_homepod_teardown}), the Azure Kinect DK camera contains a seven-microphone array~\cite{microsoft_azure_kinect_dk}, and Meta's recently announced Ray-Ban Display XR glasses have a six-microphone array~\cite{meta_neural_band}. Of course, many research systems have experimented with microphone arrays, from robots~\cite{luzanto2024effective} to smartwatches~\cite{han2017soundcraft}. Likewise, we envision our approach being used on multiple different platforms, including mobile robots, smart environment infrastructure, and worn devices with ego-centric views (see examples in Figure~\ref{fig:teaser} and Section~\ref{sec:apps}). 



As a proof of concept, we use an off-the-shelf 16-channel MiniDSP UMA-16 v2 microphone array \cite{minidsp_uma16v2}, seen in our Video Figure. The array measures 132$\times$202$\times$18~mm and enumerates as a multichannel microphone under the standard USB Audio Class (UAC) protocol. To this hardware, we add a USB webcam at the center with a 72° diagonal field of view. Both the microphone array and webcam connect to a 2024 MacBook Air M3, on which all processing (other than the VLM) occurs in real time. In Section~\ref{sec:latency}, we breakdown sources of latency.

\subsection{Conventional RGB Video and Audio Streams}
For RGB video, we use frames streamed from the USB webcam, downscaled to 640$\times$360. For audio, we create a stereo audio stream (44.1~kHz) using the upper-left and upper-right microphones on our array (12.6~cm apart). 

\subsection{Acoustic Field Video Software Pipeline}
\label{sec:software-pipeline}

Our software pipeline is agnostic to the microphone element count and geometry of the array. The latter is defined in an XML file read by our software at runtime. The sixteen-channel microphone array we used implemented the UAC protocol, allowing for plug-and-play operation. Audio was sampled at 44.1~kHz, with a chunk size of 2048, and thus our pipeline runs at $\sim$22 FPS (we discuss latency in Section~\ref{sec:latency}). 

For acoustic beamforming, we utilize the Multiple Signal Classification (MUSIC) algorithm \cite{schmidt1986music} as implemented in \textit{Acoular} Python \cite{sarradj_2025_17425756} (1,024-point FFT, Hann window, 50\% overlap). As a frequency-domain method, MUSIC requires a discrete set of frequencies for analysis. We therefore selected four center frequencies (2, 4, 6, and 8~kHz) that span a broad portion of the acoustic spectrum relevant to human activities and everyday environments. To compensate for frequency-dependent gain differences and suppress background noise, we first subtract different noise floor thresholds (18, 20, 23, and 27~dB) and then clip each SPL map to a narrow top dynamic range (0.2, 0.2, 0.5, and 0.5~dB below the maximum observed value) for our 2, 4, 6, and 8~kHz SPL maps, respectively. We then average the four resulting maps to produce a single composite acoustic field. To stabilize temporal fluctuations, we apply an eight-frame median filter (approx. 370~ms). This resulting map is rendered with a jet colormap, alpha-blended onto a gray-scaled copy of the video frame. The latter is streamed as our acoustic field video for downstream vision-language modeling, with a resolution of 640$\times$360, matching our RGB video stream.

\subsection{Zero-shot Scene Understanding}
\label{sec:VLM-AF}

We use Gemini-2.5-Pro as a representative state-of-the-art multimodal understanding model. Beyond a  language prompt, we use two input formats. First is RGB video + stereo audio, which we refer to as \textit{Conventional Video}. Second, we have the RGB video + acoustic field video, arranged as a stacked pair, plus stereo audio, which we refer to as \textit{Conventional + Acoustic Field Video.}

We interface with Gemini using the google-generativeai Python API \cite{google-generativeai}. We use the following prompt:

\begin{PromptText}
Be definitive in your answers. Avoid hedging words like "potentially", "possibly", or "probably", and other speculative language. Also, answer concisely; one or a few sentences at most. I am giving you a video clip with audio of a scene.
\end{PromptText}
\vspace{-0.28em}
\begin{PromptTextBlue}
The video clip has two synchronized visualizations of the same camera view. The top is the standard video of the scene. Bottom is the same video, but overlaid with a sound pressure map (jet color scheme, e.g., blue is no or low sound, and oranges and reds are louder sound sources). The sound pressure map shows where sounds are coming from. Your answer should not explicitly mention the video or the sound pressure map.
\end{PromptTextBlue}
\vspace{-0.28em}
\begin{PromptText}
Using this information, I want you to answer the following question:
\end{PromptText}

\hfill

\noindent Only the black portions of the prompt are used when inputting \textit{Conventional Video} (for baseline evaluation purposes, discussed in Section~\ref{sec:eval}), and the complete prompt (black and blue portions) is used when inputting \textit{Conventional + Acoustic Field Video}. 

\subsection{Live Mode}
We also take advantage of Gemini 2.5 Pro's ``Talk to Gemini Live" mode to ingest a live audio-video stream, which could come from a robot, AI glasses, etc. In our case, we pass Gemini our \textit{Conventional + Acoustic Field Video}. We use the same prompt as the prior section, sans the very last sentence. In this way, it is open-ended for the user to define a desired assistive use. They could say, for instance, \textit{``let me know if there are any hazards while I am working"}, or \textit{``keep me informed about the status of my 3D print job"}. We offer some illustrative examples in our Video Figure. 

\subsection{Latency}
\label{sec:latency}
Our approach is comparatively lightweight with modest latency; VLM token generation dominates any interaction. There are two main sources of latency in our system: sensor latency and beamforming computation. Starting first with sensors: the USB Audio Class (UAC) protocol is highly optimized on modern operating systems, with just a few milliseconds of latency, while USB Video Class (UVC) devices have on the order of tens of milliseconds of latency. Our beamforming stage (with four runs of the MUSIC algorithm) takes 58~ms on our 2024 MacBook Air M3. These two main processes, plus various other smaller latencies (e.g., moving video/audio frames into Python, rendering the acoustic field onto live video for debugging) combine for a total latency of around 90~ms. We note our Python-based pipeline is not heavily optimized, and this reported latency should be considered an upper bound.

\begin{figure*}
    \centering
    \includegraphics[width=\linewidth]{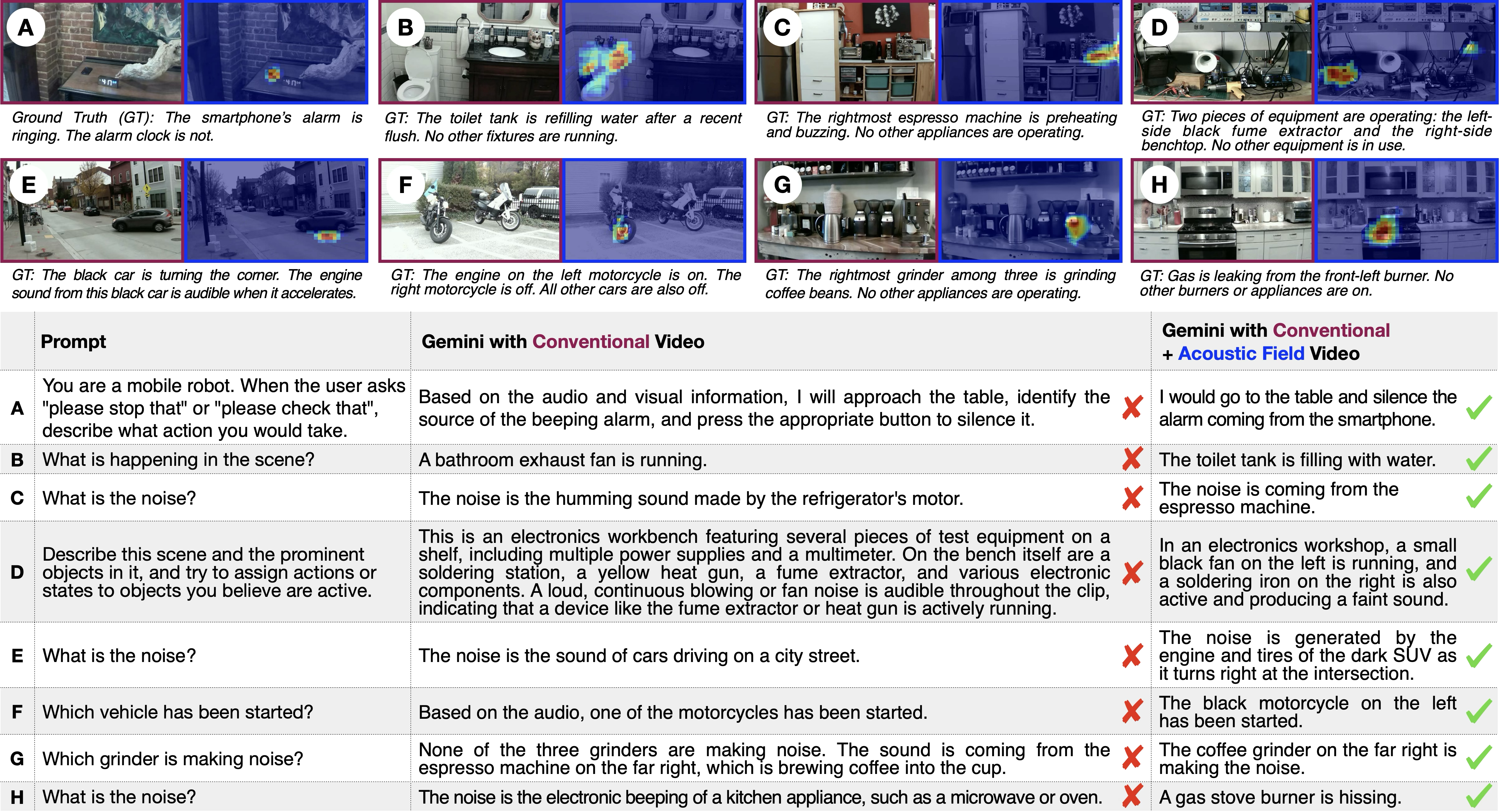}
    \caption{Top: Example scenes (conventional RGB and acoustic field still frames shown) drawn from our test set. Bottom table: Example prompts, along with output from Gemini given either \textit{Conventional Video} or \textit{Conventional + Acoustic Field Video} as input. }
    \label{fig:successful_examples}
\end{figure*}

\section{Potential Applications}
\label{sec:apps}
We envision AI agents augmented with acoustic field sensing being valuable across a wide range of scenarios. Below we outline several representative use cases that illustrate the potential of this modality to enhance perception, interaction, and situational awareness. Please also see our Video Figure for additional examples using our live-mode implementation. 

\subsection{Worn Devices (Ego-Centric View)}
Worn devices such as AI pins and XR glasses have a unique ego-centric view onto the world. With acoustic field video, such systems could infer where sounds originate relative to the user, allowing them to better understand and monitor activities, interactions, and context. For example, AI glasses could proactively let users know “the kettle is boiling” (Figure~\ref{fig:teaser} middle), ``you left the stove on" (Figure~\ref{fig:successful_examples} H), and ``you should turn on the fume extractor while soldering" (Figure~\ref{fig:failure_attention} I). 

\subsection{Mobile Robots}
For autonomous mobile robots, spatialized acoustic understanding enhances both perception and interaction. Robots equipped with microphone arrays can identify and localize sounds that indicate human presence, mechanical operation, or environmental changes. For instance, a household robot could infer that a laundry machine has finished its cycle, or that someone called its name from another room (Figure~\ref{fig:teaser} bottom). In workplace settings, acoustic cues such as machine vibrations, dripping taps, alarms, can reveal operational states that are difficult to capture visually (Figure~\ref{fig:teaser} \& \ref{fig:successful_examples}). 

\subsection{Smart Speakers \& Smart Environments}
Although fixed in position, smart speakers, security systems, and other ambient IoT infrastructure occupy ideal vantage points for continuous monitoring of activity in homes and workplaces. When augmented with vision and acoustic field sensing, these systems could move beyond simple sound detection (e.g., breaking glass alert) to spatial reasoning (e.g., user accidentally dropped a glass). For example, a smart speaker could localize a crying infant, detect if a stove burner was left on (Figure~\ref{fig:successful_examples} H), track the usage of appliances (Figure~\ref{fig:teaser} middle), monitor water consumption (Figure~\ref{fig:successful_examples} B), or identify noise from a kitchen appliance (Figure~\ref{fig:failure_attention} K). In office or factory environments, acoustic fields could integrate with building systems such as HVAC or lighting to adapt to occupancy and activity levels. Moreover, these maps could inform nearby robots or embodied agents, allowing a cooperative ecosystem where devices share a common, spatially grounded understanding of sound events in their surroundings.

\section{Evaluation}
\label{sec:eval}

\subsection{QA Scene Data Collection}
There is no public data set that contains paired conventional video and acoustic field video. It may be possible to partially simulate such data using audiovisual segmentation~\cite{senocak2024aligning, Sun_2023_CVPR}, however results can be inconsistent if there are multiple objects of the same class in the scene or if sounds are generic (hums, vibrations, whirs, electronic beeping, etc.) and could be attributed to more than one present object. For this reason, we do not believe there is a suitable substitute for real-world data at this time, and as such, we created our own dataset for evaluation, which we also make freely available for replication and advancement. 

All recordings were captured using the hardware and software described in Section \ref{sec:implementation}. One capture instance consists of a synchronized, five-second triplet of data: conventional (RGB) video, acoustic field video, and stereo audio. Data collection was conducted in ten diverse environments: bathroom, bedroom, kitchen, office, office kitchenette, utility room, electronics room, fabrication workshop, parking lot, and a road. We did not control for background noise, and so most of our data contains, e.g., HVAC noise and background chatter.

We endeavored to curate a diverse range of questions that would be applicable to different use cases (see Section~\ref{sec:apps}). For all instances captured, we asked three common questions: First, simulating a mobile robot use case, we asked: ``When the user says `please stop that' or `please check that', what action would you take?". Targeting more generic scene understanding, we also asked: ``What is happening in the scene?", and ``What is the noise?". In addition to these three common questions, we also included at least one (max five) custom scene-specific queries, such as in the office kitchenette: ``What stage of coffee making are we in?". In total, our evaluation set contains 402 QA instances.

\subsection{Procedure}
To test whether adding acoustic field video improves zero-shot scene understanding, we compare two input modalities: \textit{Conventional Video} (i.e., RGB video + stereo audio; baseline) vs. \textit{Conventional + Acoustic Field Video} (ours). The task is to answer a question for a given five-second question-scene instance. We use the same model, Gemini 2.5 Pro, for both conditions (see prompt in Section~\ref{sec:VLM-AF}). To prevent context carryover and standardize quality, we initialize a new inference session for each instance. As we have 402 instances, tested with two input modalities, our procedure generated 804 QA instance pairs. Including uploading data to Gemini, this automated process took around 3 hours.

\subsection{Human Raters}
To evaluate the quality of VLM output, we recruited three human raters (mean age~=~20.3; two identified as women). We developed a basic web interface to facilitating process all of the data. For each QA instance, raters (wearing headphones) watched the five second conventional (RGB) video clip including stereo audio, along with a short ground-truth text description and question about the scene. Raters could re-play the video as they saw fit to understand each scene and question being asked. 

Once satisfied they understood the scene, the raters proceeded to evaluate the correctness of a VLM-generated answer. Two buttons were offered: "correct" and "not exactly". The raters were told during orientation that an answer "should only be marked 'correct' if it matches the ground truth without contradictions, omissions, or vague language," while answers with "with wrong or vague responses should be marked as 'not exactly'." Upon selecting one of these options, they were shown a second VLM-generated answer, and again told to assess correctness. The presentation order of the two input modality conditions was randomized and counterbalanced. Lastly, for the same QA instance, raters were shown both answers side-by-side and asked: "Which do you believe is the better answer?". They could select the first answer, second answer, or equal preference. For final labels, we took the raters' majority vote; i.e., for an answer to be deemed correct, incorrect, or preferred, two of three or three of three raters had to rate it as such. There is a potential corner case with this methodology, which is a three-way tie between preference for answer A, answer B, and equal preference, but this never occurred in our data.


The raters completed this task independently, which took around 4 hours to complete. In total, our raters provided 3618 responses (402 QA instances $\times$ 3 ratings $\times$ raters). We see substantial agreement on correctness for both input modalities: Fleiss’ $\kappa=0.72$ for \textit{Conventional Video} and $\kappa=0.65$ for \textit{Conventional + Acoustic Field Video}. When our raters had a preference between the two answers, we see a similarly high inter-rater reliability, with a Krippendorff’s $\alpha$ of 0.78.



\section{Results \& Discussion}
\begin{figure}
    \centering
    \includegraphics[width=1\linewidth]{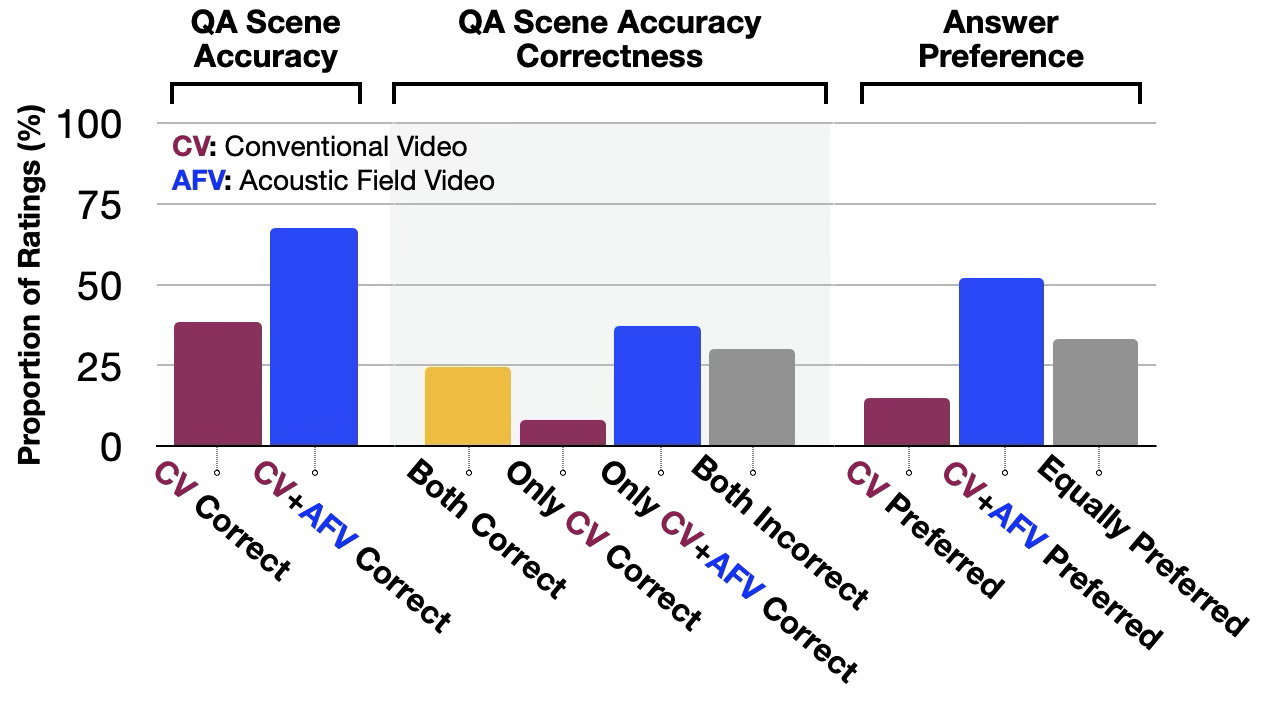}
    \vspace{-5mm}
    \caption{QA scene understanding with and without acoustic field video. Left: Overall answer accuracy when the VLM sees only \textit{Conventional Video} (CV) versus \textit{Conventional + Acoustic Field Video} (CV+AFV); Middle: Breakdown of correctness. Right: Human raters’ answer preferences.}
    \label{fig:results}
\end{figure}

We now present our main findings, also summarized in Figure~\ref{fig:results}, before moving to the discussion.

\subsection{QA Scene Accuracy}
\label{sec:QAAccuracy}
Across all 402 QA instances, the baseline \textit{Conventional Video} input condition was rated as 38.3\% correct, while our \textit{Conventional + Acoustic Field Video} input condition was rated as 67.4\% correct. The magnitude of this increase strongly suggests that acoustic field data materially helps the VLM disambiguate sound sources and relate audio evidence to the visual scene. 

Breaking down this data a different way, we see that in 24.4\% of QA instances, the raters believed both VLM input conditions were incorrect, and 30.1\% of the time raters believed both input conditions were correct. More interesting is when answers derived from \textit{Conventional Video} were rated as correct, while answers from the \textit{Conventional + Acoustic Field Video} were rated as incorrect (8.4\% of QA instances). However, a much larger proportion shows the opposite behavior, where answers from the VLM given \textit{Conventional + Acoustic Field Video} were rated as correct while the \textit{Conventiona Video} input was wrong (37.3\%). 


\subsection{Answer Preference}
When comparing answers from the two VLM input conditions side-by-side, answers derived from \textit{Conventional Video} input were preferred in 14.9\% of the time, while answers from \textit{Conventional + Acoustic Field Video} input were preferred 52.0\% of the time (with the remaining 33.1\% being judged as equally valid). 

\subsection{Improved Attention}
We observed that when an acoustic field video is provided, the VLMs attention improved, leading to both more correct and more succinct answers (see examples in Figure~\ref{fig:failure_attention}, I-M). The VLM also tended to elevate the active object to the beginning of its reply. Lastly, we also saw the absence of sound in the acoustic field being utilized. For example, as can be seen in Figure~\ref{fig:failure_attention} I, where a user is soldering without an exhaust fan running, the VLM is able to reason that the fan is off using the acoustic field.





\subsection{Failure Cases}
\begin{figure*}
    \centering
    \includegraphics[width=\linewidth]{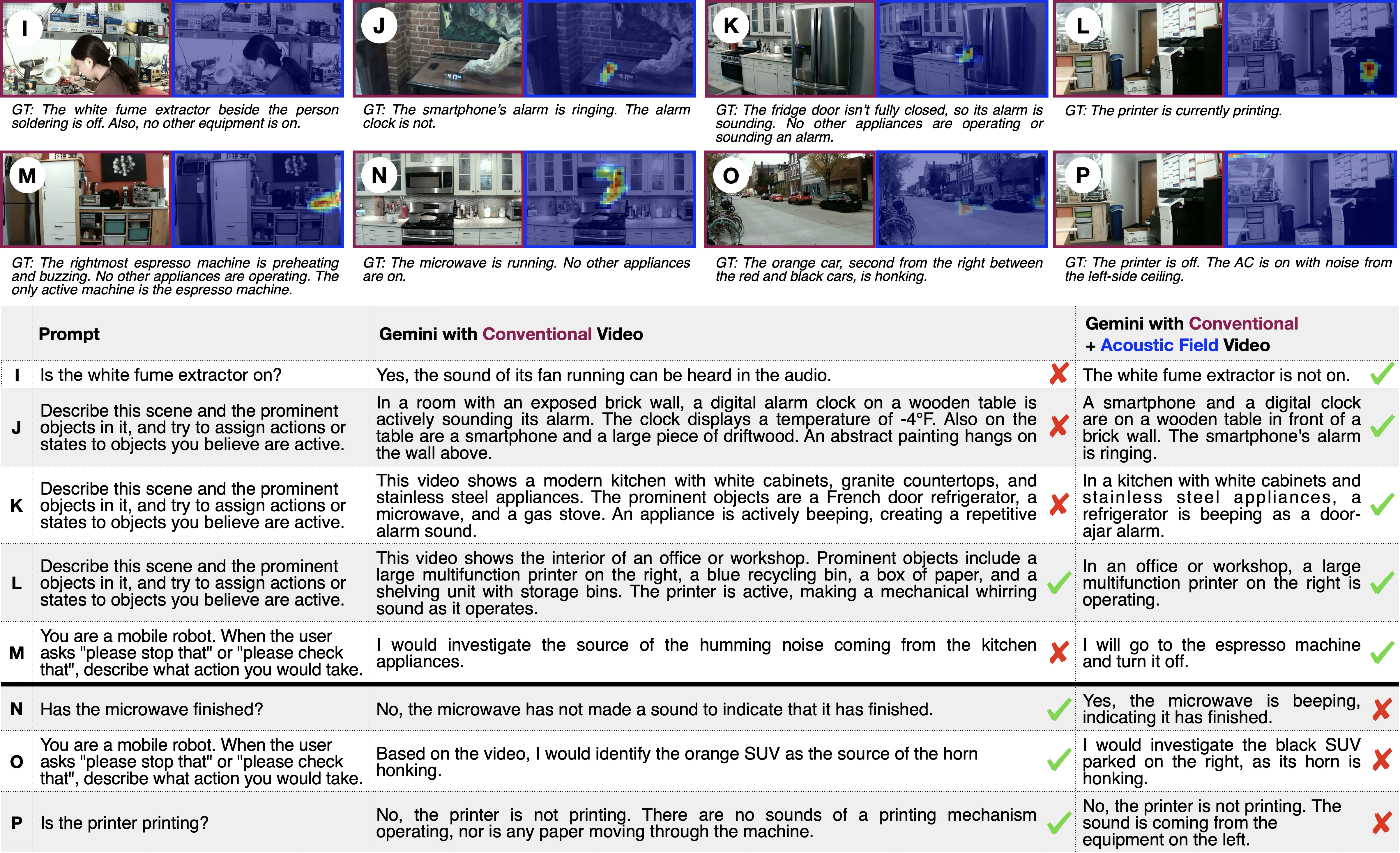}
    \caption{Examples of good attention and failure cases (separated by black horizontal rule). Top: Example scenes (\textit{Conventional + Acoustic Field Video}) drawn from our test set. Bottom table: Example prompts, along with output from Gemini given either \textit{Conventional Video} or \textit{Conventional + Acoustic Field Video} as input.}
    \label{fig:failure_attention}
\end{figure*}

We also noted some interesting failure cases, a few of which we highlight in Figure~\ref{fig:failure_attention}, N-P. As reported in Section \ref{sec:QAAccuracy}, only in 8.2\% of the QA instances did the VLM get the correct answer using \textit{Conventional Video}, but got the wrong answer using \textit{Conventional + Acoustic Field Video}. We analyzed these 33 failure instances for high-level themes. We found occasional misclassifications of sounds --- for instance, the humming of a running microwave was described as "beeping" and "finishing". We also observed instances of misattributions of sound --- for example, at the beginning of one clip, a parked car honks its horn which is clearly visible in the acoustic field video; however, another car drives through the scene in the second half of the clip (which is very salient) and the honk is incorrectly attributed to the moving car. Finally, we saw in some clips with quiet but audible background noise (especially from HVAC) that the model sometimes attributed activity to inactive objects in the scene, even without a cue present in the acoustic field (e.g., an inactive 3D printer sitting on a table). It may be that prompt engineering could resolve some of these issues. 

\section{Limitations \& Future Work}
While our results demonstrate clear benefits of incorporating acoustic field video into multimodal reasoning, there are several limitations of note. First, our evaluation dataset, although diverse in scene type and acoustic configuration, remains modest in size and limited to real-world recordings from a single array geometry. Broader datasets spanning varied microphone arrangements, room impulse characteristics, more outdoor environments, and cluttered multi-source scenes will be important to generalize our initial findings.

We also note that our implementation uses frequency-domain MUSIC beamforming with a small set of discrete analysis frequencies. Although this approach offers high angular resolution and robustness, it imposes computational overhead and a fixed spatial grid that may limit responsiveness in dynamic scenes. Moreover, MUSIC relies on narrowband assumptions and is known to degrade in low-SNR settings, highly reverberant environments, or when sound sources exhibit broad spectral content not well captured by our chosen frequency bands. Future work could explore alternative wideband beamforming techniques, neural beamformers, learned spatial acoustic encoders, or hybrid classical–neural architectures that jointly optimize spatial and semantic representations. Nonetheless, MUSIC was sufficient for our proof-of-concept implementation (i.e., the goal of this work was not to make advances in beamforming, but rather explore the potential utility of the signal). More advanced methods would likely unlock further accuracy gains against conventional video inputs. 

Additionally, while acoustic field video provides explicit spatial grounding, VLMs ingest it only through conventional visual pathways. As a result, the model performs no audio–acoustic fusion at the level of raw spatial features; instead, spatial sound structure must be inferred from a colorized overlay. Training VLMs (or dedicated multimodal encoders) to directly process acoustic fields in a native tensor format may unlock significant performance gains and reduce reliance on handcrafted visualization choices (e.g., colormaps, clipping thresholds). Similarly, joint training on synchronized RGB, audio, and acoustic field data could yield richer cross-modal bindings and more consistent attribution of sounds to objects.

Finally, although we evaluate zero-shot performance, in-the-wild deployment introduces additional challenges: user motion with wearable devices, array self-noise, user-motion generated noise, and device heterogeneity within and across manufacturers. Understanding the sensitivity of acoustic-field-driven reasoning to these factors, and developing calibration-free or self-adaptive techniques, remains an open question for deployment. 

\section{Conclusion}
We presented acoustic field video as a new spatially grounded acoustic modality for multimodal vision–language reasoning. By visualizing where sound originates and aligning this information with RGB video and audio, acoustic field video enables VLMs to connect auditory evidence to specific objects and locations in a scene. Using a real-time, low-cost beamforming pipeline and a 402 QA-instance benchmark, we show that augmenting conventional RGB+audio inputs with acoustic field video yields substantial gains in zero-shot scene understanding and is consistently preferred by human raters. As microphone arrays continue to proliferate across smart speakers, robots, and wearable devices, our results highlight acoustic field video as a practical and powerful path toward more perceptually grounded multimodal intelligence.


{
    \small
    \bibliographystyle{ieeenat_fullname}
    \bibliography{main}
}


\end{document}